\documentclass[prb,twocolumn,showpacs,preprintnumbers,amsmath,amssymb]{revtex4}
\usepackage{graphicx}
\usepackage{dcolumn}
\usepackage{bm}

\begin{document}

\preprint{APS/Manuscript}

\title{ The Structural Stability and Energetics
of Single Walled Carbon Nanotubes Under Uniaxial Strain}

\author{ G. Dereli}
\affiliation {Department of Physics, Middle East Technical University, 06531 Ankara, Turkey}
\email {gdereli@metu.edu.tr}
\author{C. \"{O}zdo\u{g}an}
\affiliation {Department of Computer Engineering, \c{C}ankaya University, 06530 Ankara, Turkey  }
\email{ ozdogan@cankaya.edu.tr}

\date{\today}

\begin{abstract}
$(10 \times 10)$ SWNT consisting of 400 atoms with 20 layers is
simulated under tensile loading using our developed $O(N)$
parallel TBMD algorithms. It is observed that the simulated carbon
nanotube is able to carry the strain up to 122\% of relaxed tube
length in elongation and up to 93\% for compression. The Young's
modulus, tensile strength and Poisson ratio are calculated and the
values found are 0.311 TPa, 4.92 GPa and 0.287, respectively.
Stress-strain curve is obtained. The elastic limit is observed at
the strain rate of 0.09 while the breaking point is at 0.23. The
frequency of vibration for the pristine $(10 \times 10)$ carbon
nanotube in radial direction is $4.71 \times 10^3$ GHz and it is
sensitive to the strain rate.
\end{abstract}
\maketitle


Single walled carbon nanotubes (SWNT) are observed to have high flexibility, strength and stiffness, very similar to those of individual graphene sheets.
Eventhough direct measurements of such mechanical properties are difficult to perform
due to the nanosizes involved; along with the developments in instrumentation, production, processing and manipulation techniques, the measurements of the elastic moduli of carbon nanotubes has become possible. On the other hand,  this extreme small size is very suitable for performing atomistic simulations.
Both the experimental \cite{xilipa}--\cite{yufiar} and theoretical \cite{yazhch}--\cite{yacabr} studies have shown
that SWNTs, and SWNT ropes are promising low--weight, high--strength fibers for use as reinforcing element in
composite materials. It is also predicted that SWNT's can sustain large strains in the axial direction \cite{yacabr}.
The axial Young's modulus values range from 200 GPa to 5.5 TPa in literature.

\medskip

The carbon nanotubes are known to  have honeycomb structure \cite{sadrdr} and three C--C bonds formed
between a carbon atom and three adjacent carbon atoms in the unit cell can be classified as two kinds
according to the relations of their spatial orientations; one is perpendicular to tubular axis and the
other is not perpendicular to tubular axis in armchair tube.
In this study,  an armchair  $(10 \times 10)$ carbon nanotube consisting of 400 atoms with 20 layers
is simulated
under tensile loading  using our developed O(N) parallel TBMD program
\cite{ozdeca},\cite{deoz}. Two steps are followed; firstly, the tube is annealed at simulation
temperature for 3000 MD steps (each time step used in the simulation being 1 fs). The variation of total
energy  and some physical properties such as radial distribution function, atomic coordination number,
 bond--angle distribution function, and bond--length distribution function are given.
 The second step is to apply strain on the tube. Several  groups have proposed different procedures; such as shifting the end atoms along the axis (i.e. z--direction) by small steps \cite{oziwmi}, reducing radial dimension while the nanotube is axially elongated \cite{zhdugu}, pulling in the axial direction with a prescribed strain rate and following each step of pulling by some additional MD steps in order to relax the distorted structure \cite{zhoshi} and finally using hydrostatic pressure exerted on the walls of SWNT by the means of encapsulating $H_2$ molecules inside the tube and the wall of the tube \cite{xizhma}. In our study, the tensile strain is applied by the reduction or enlargement of the radial dimension while the nanotube is axially elongated or contracted. Throughout this procedure  volume of the tube is kept constant. Zhou et al. \cite{zhdugu}, has investigated the mechanical properties of SWNT with the same procedure  using a first--principle cluster method within the framework of local density approximation. We further simulated the deformed tube structure (under uniaxial strain) for another 2000 MD steps (time step is chosen again as 1 fs) to understand the strain mechanism. Strain is obtained from $\varepsilon=\left( {L-L_0 \over L_0}\right) $, where $L_0$ and $L$ are the tube lengths before and after the strain, respectively. Several strain values are applied to a pristine tube to study the strain rate.  Simulations are performed at room temperature and periodic boundary condition is applied  along the axial direction.

\medskip

\begin{figure}
\vskip 18 cm \includegraphics{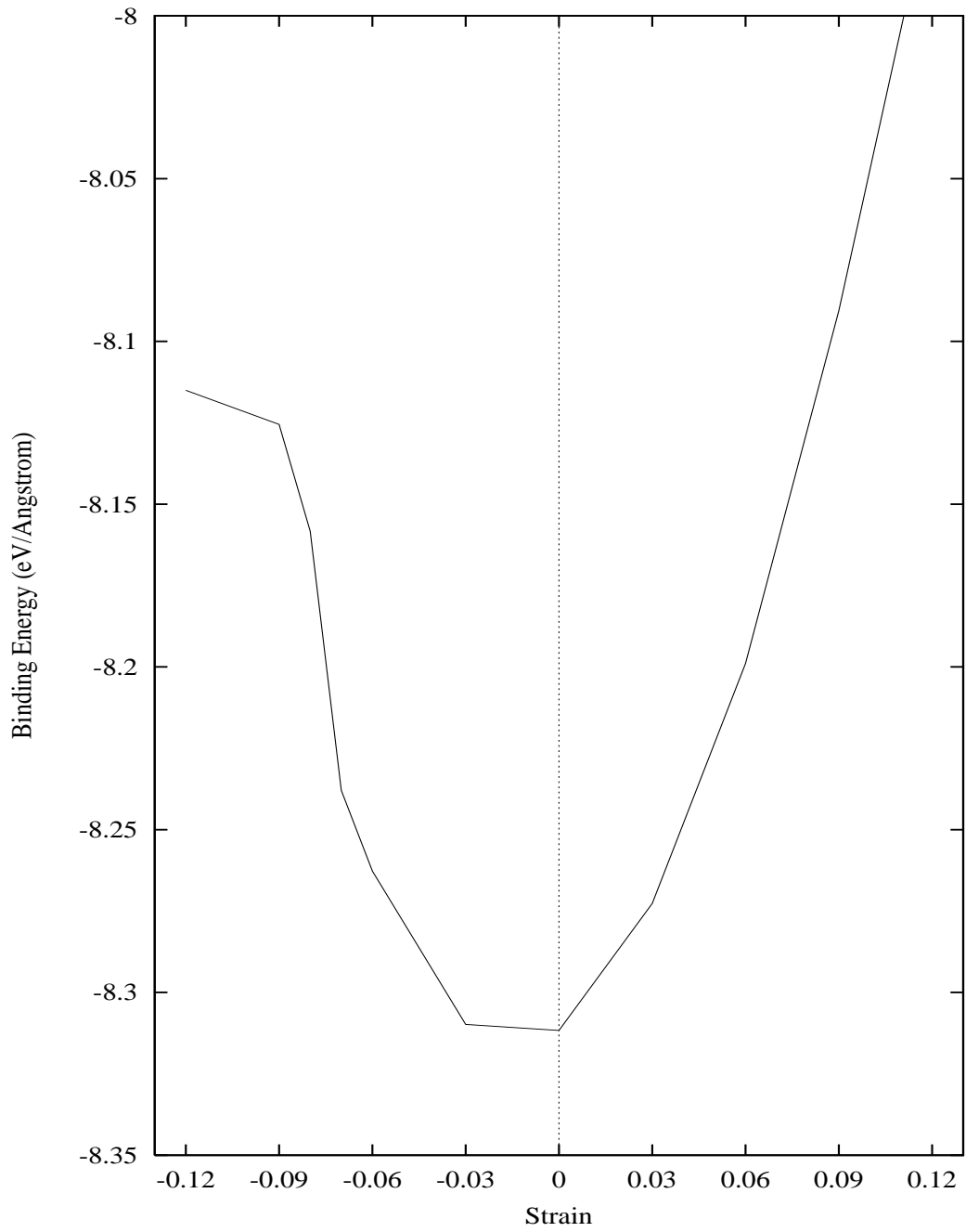} \vskip 1.5 cm
\caption{Total energy curve as a function of strain $\epsilon$}
\label{4.4.2}
\end{figure}

The binding energy  curve of a carbon nanotube as a function of
the strain along the tubular axis is given in  Figure \ref{4.4.2}.
We observe an asymmetric pattern for the cases of elongation and
compression. The tube does not have a high strength for the
compression as much as for elongation. This might be due to the
dominant behavior of repulsive forces in the system under uniaxial
strain. This figure indicates that the remarkable elastic
properties under large strains are caused by nonparabolic strain
energy. We observed  that the carbon nanotube is able to carry the
strain up to 122\% of the pristine tube length in elongation and
up to 93\% of pristine tube length in compression.

\begin{figure}
\vskip 18 cm \includegraphics{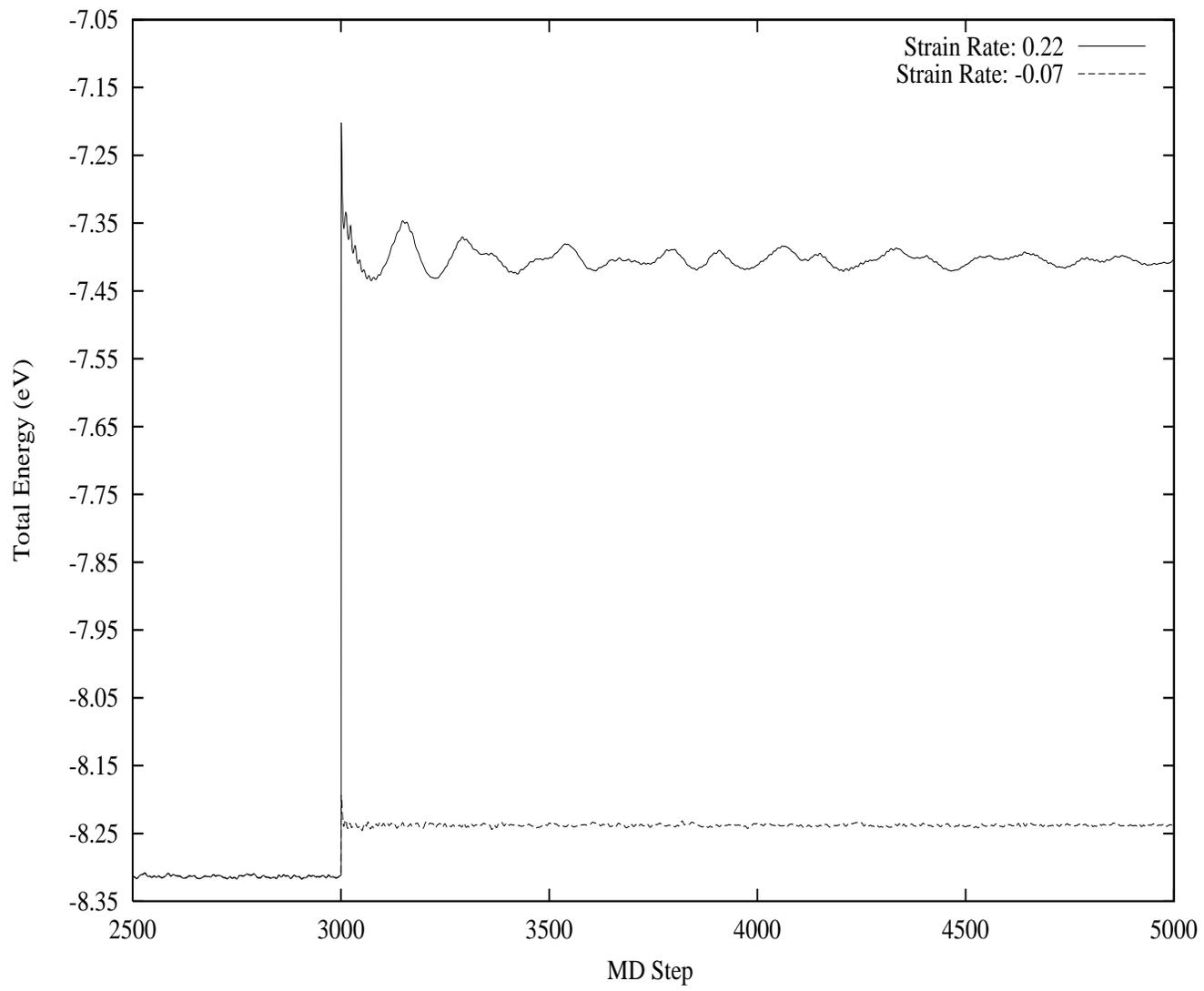} \vskip 1.5 cm
\caption{The variation of total energy of deformed 10x10 Carbon
nanotube during MD simulation for the strains 0.22 and -0.07
(negative sign corresponds for the compression), respectively.}
\label{4.4.3}
\end{figure}

\begin{figure}
 \vskip 18.0 cm
    \includegraphics{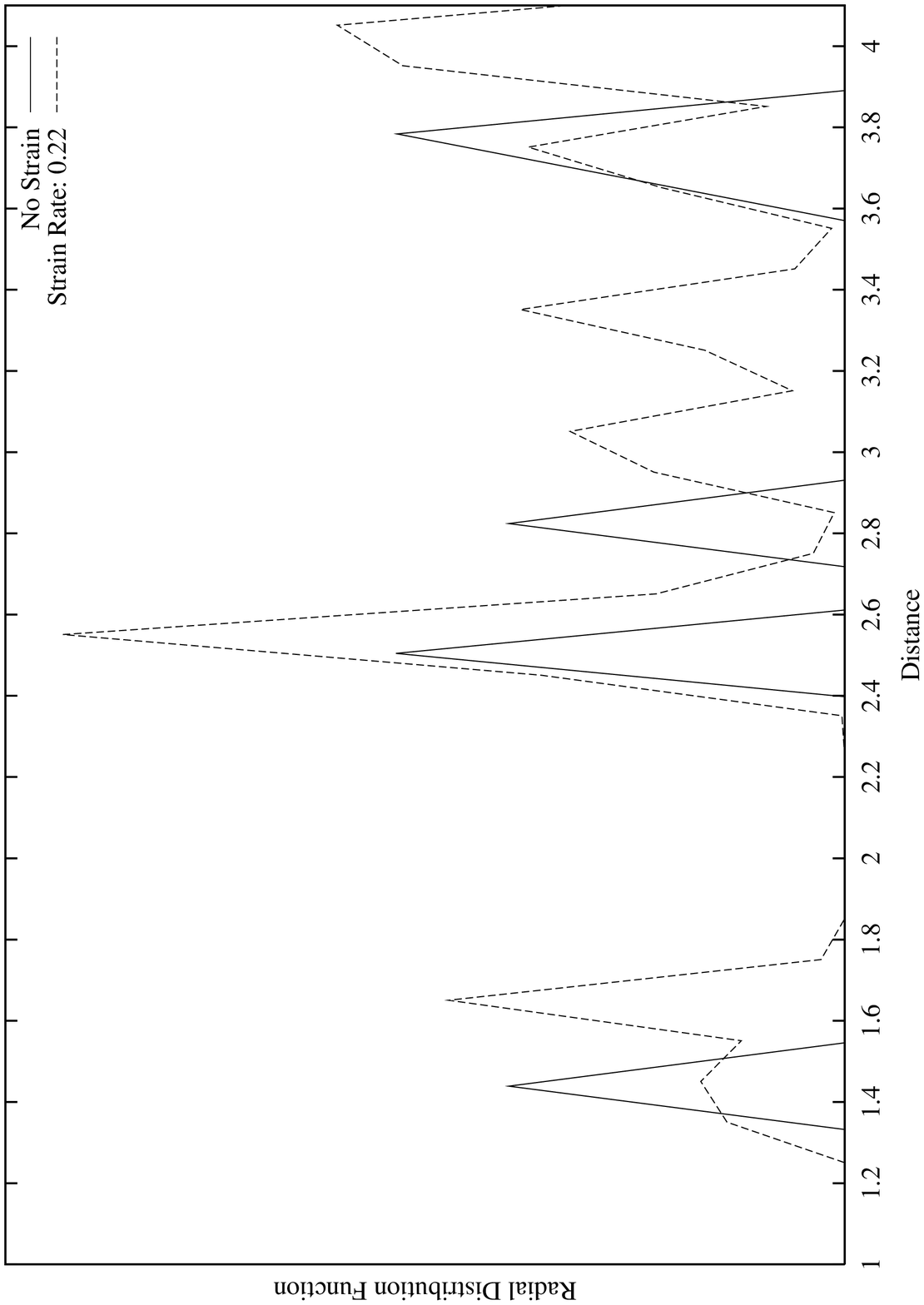}
\vskip -18.0 cm $\left. \right.$
\vspace{2.5cm}

\vskip 18.0 cm
    \includegraphics{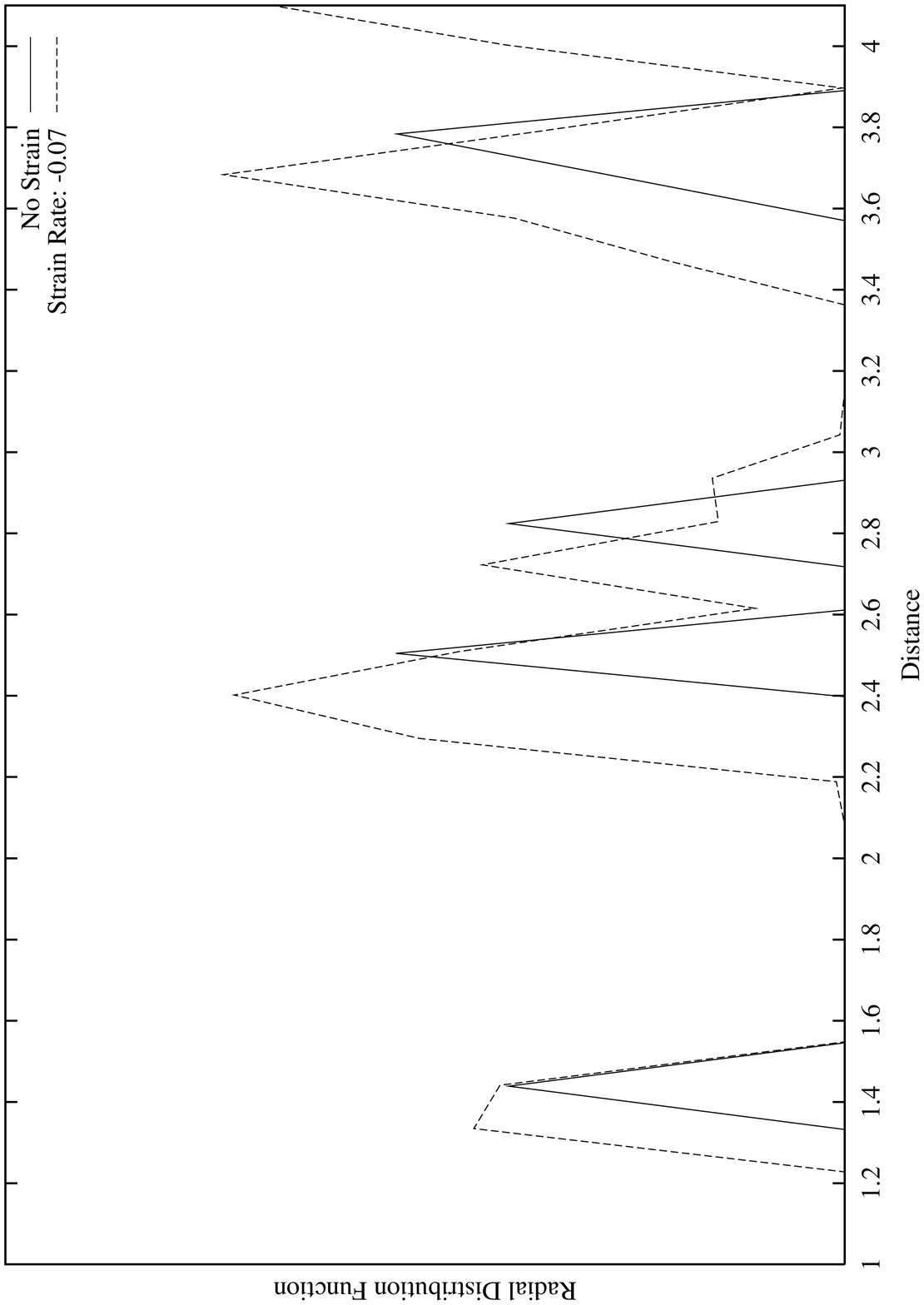}
\vskip 1 cm
\caption{Radial Distribution Functions for the Tube Structure 10x10 under strains 0.22 and -0.07, respectively.}
\vspace{2.5 cm}
\label{4.4.6}
\end{figure}

\begin{figure}
\vskip 18 cm \includegraphics{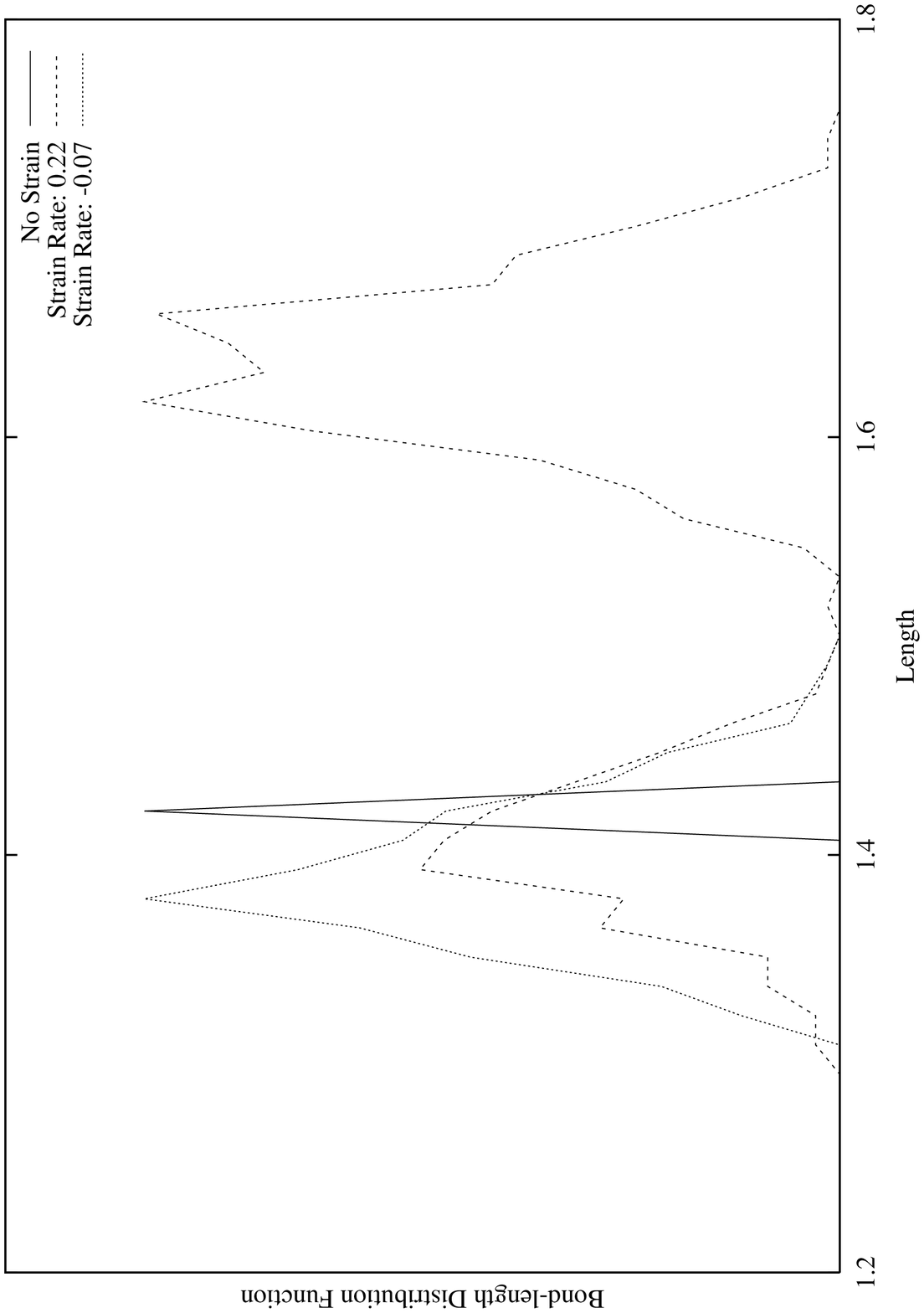} \vskip 1.5 cm
\caption{Bond--length Distribution Functions for the Tube
Structure 10x10 under strains 0.22 and -0.07, respectively.}
\label{4.4.7}
\end{figure}

\begin{figure}
\vskip 18 cm \includegraphics{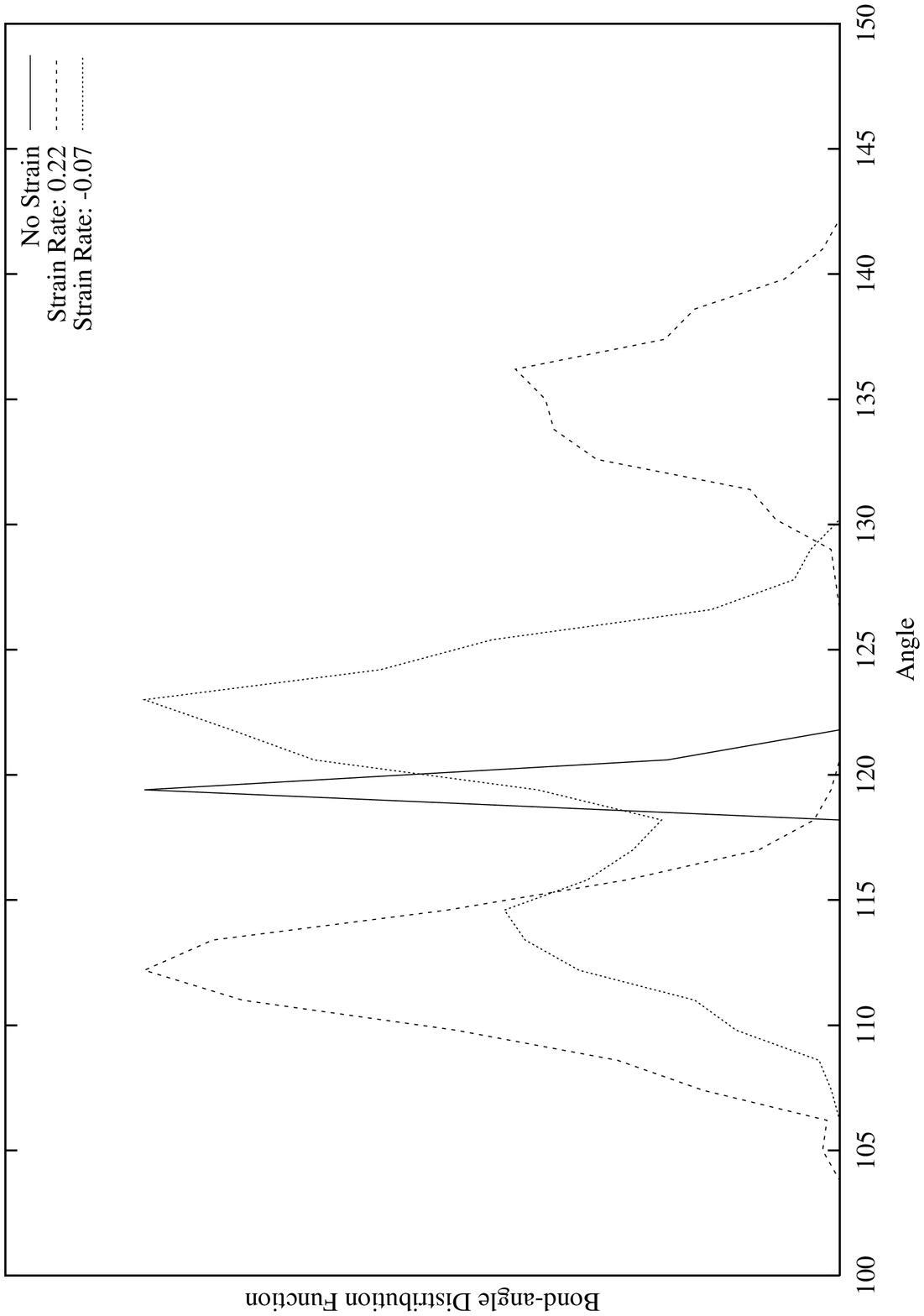} \vskip 1.5 cm
\caption{Bond--angle Distribution Functions for the Tube Structure
10x10 under strains 0.22 and -0.07, respectively.} \label{4.4.8}
\end{figure}

The variation of total energy of the deformed system during MD simulation for the strains of 0.22 in
elongation and 0.07 in compression are given in the Figure \ref{4.4.3}. In the graphs, first 3000 MD
steps is for the equilibration of the carbon nanotube and the next 2000 MD steps shows the variation of
 the total energy of the tube during the simulation under the applied uniaxial strain. It is seen that
 the tube under these strain rates is able to sustain its structural
 stability.For high strain rates the changes in the radial distribution function, bond--length distribution and bond--angle distribution are given in Figs. \ref{4.4.6}--\ref{4.4.8}. It is seen in the graphs that the two--third of the bond lengths and one--third of the bond angles increases (decreases) when tube is elongated (contracted), as expected.
  Bond angles and bond lengths are the two important factors that control the deformation.
 The effect of the strain on the bonds is such that it  alters the angles between two neighboring carbon
 bonds and changes the lengths of the C--C bonds.

\begin{figure}
\vskip 18 cm \includegraphics{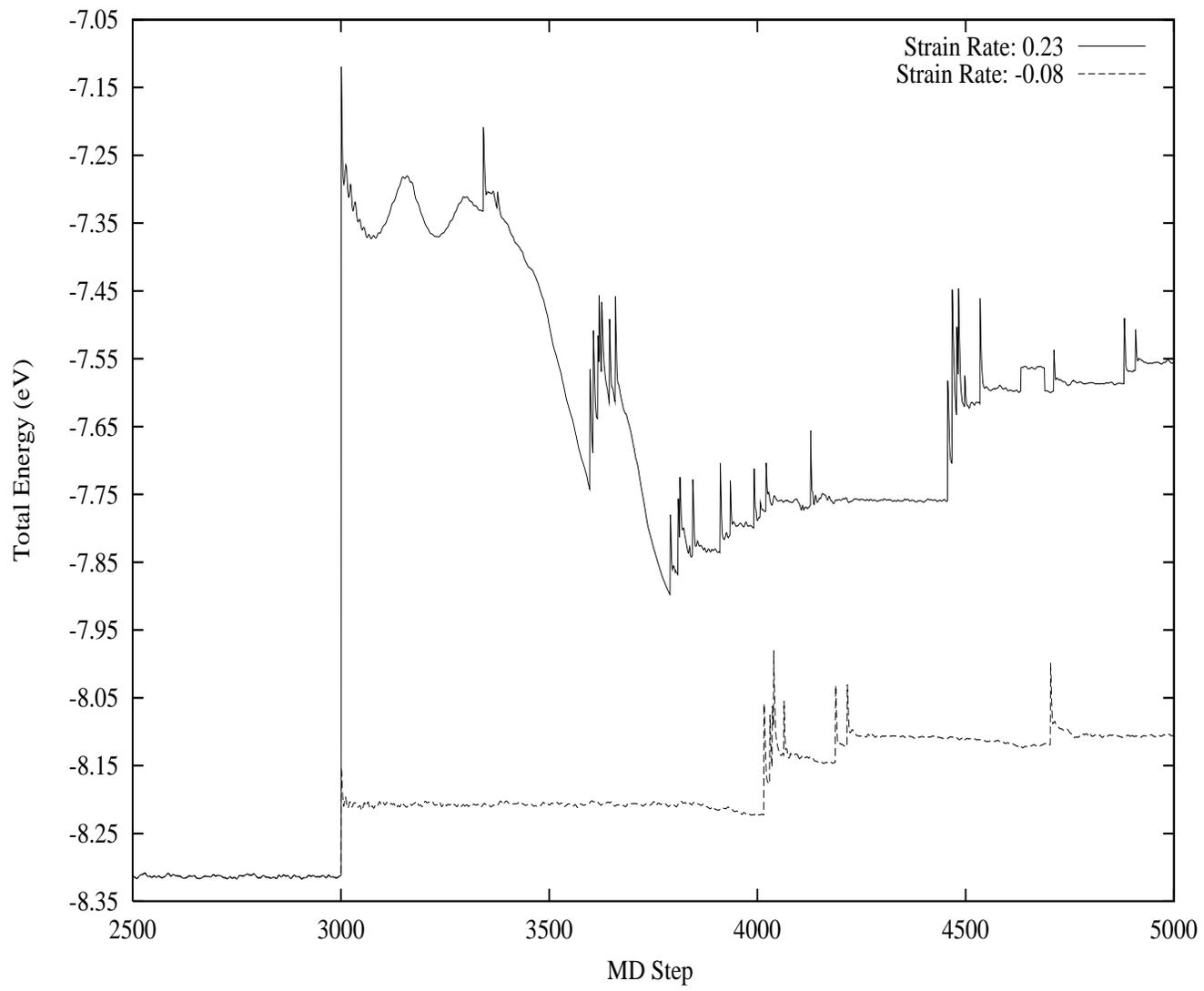} \vskip 1.5 cm
\caption{The variation of total energy of deformed 10x10 Carbon
nanotube during MD simulation for the strains 0.23 and -0.08,
respectively.} \label{4.4.11}
\end{figure}

Increasing the strain beyond these points results in
disintegration of atoms from the carbon nanotube. The geometrical
structures, and the behavior of the total energy for the strains
0.23 (elongation) and 0.08 (compression) show  that the elongated
tube dissociates by starting from the middle like a zipper while
the compressed tube starts to dissociate from the ends of the
tube. Each peak in Figure \ref{4.4.11}
represent disintegrations of the atoms from the tube.

\begin{figure}
\vskip 18 cm \includegraphics{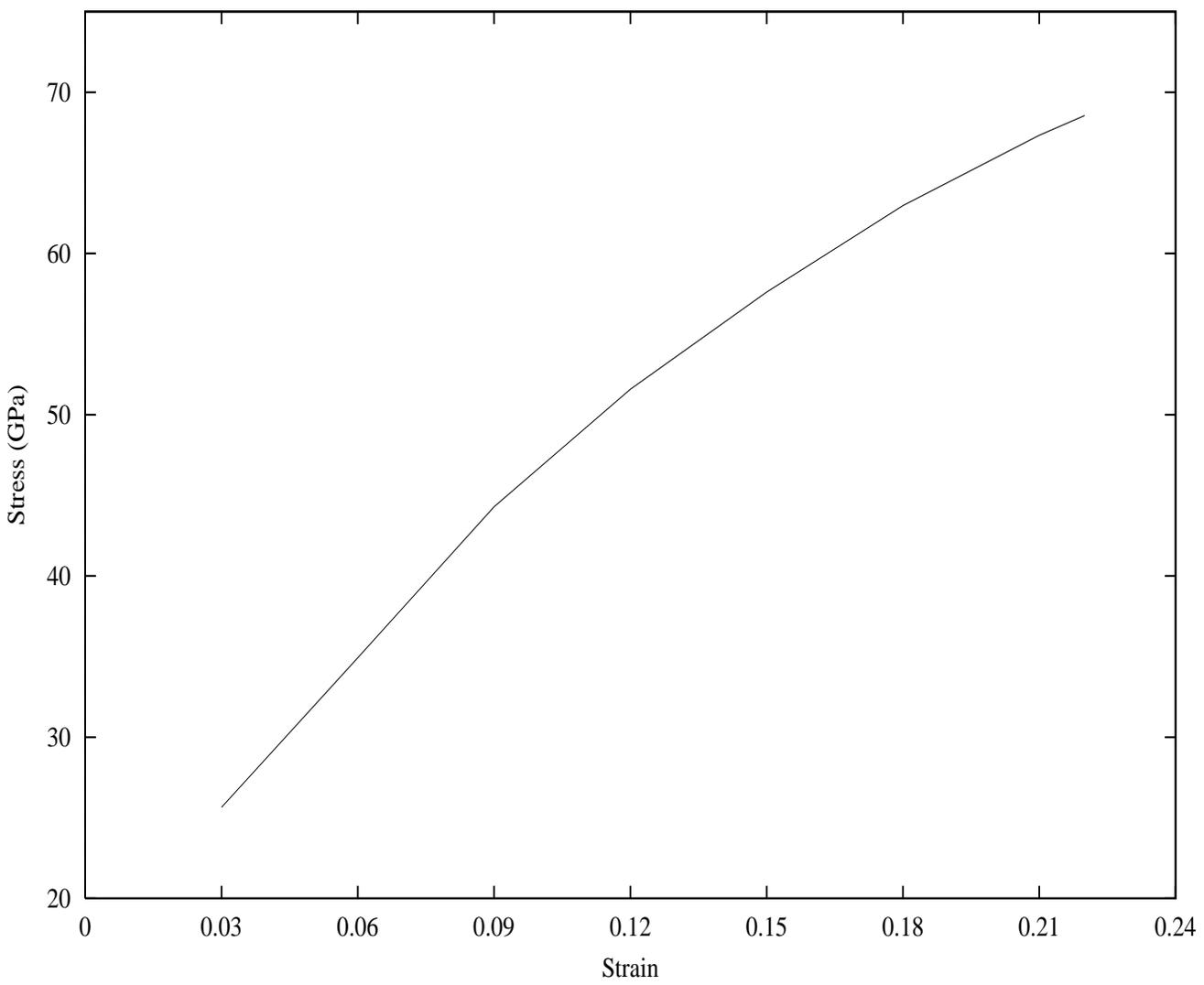} \vskip 1.5 cm
\caption{The uniaxial stress applied to the tube versus the strain
$\epsilon$ (elongation).} \label{4.4.12}
\end{figure}

The elastic constants are calculated from the second derivative of the energy density with respect to various strains. To obtain the stress--strain curve, the cross--section upon which the resulting forces act is needed to be estimated. The cross--sectional area of a nanotube is ambiguous in definition \cite{sadrdr}. If a circular cylindrical shell is considered around the surface of the nanotube, then the surface area of the cross--section $s$, is defined by
\begin{equation}
s=2\pi R\delta R
\end{equation}
where $R$ stands for the radius of the SWNT and $\delta R$ for the wall--thickness. It should be noted that different wall--thickness values were used by several groups
\cite{yabrbe}--\cite{lialdo}.
In Ref. \cite{lu}, $\delta R$=3.4 \AA ~(measured interwall distance in the Multi Wall Nano Tube) was used, while in Ref. \cite{lialdo} $\delta R$=1.7 \AA~ (taken as the van der Waals radius for Carbon) was accepted , it is also accepted as 0.66 \AA~ (in the $\pi$ orbital extension) in Ref. \cite{yabrbe} and as the whole cross-sectional area of the tube in the Ref. \cite{corwil}. We defined the thickness of SWNT shell as $\delta R=3.4~$ \AA. The stress--strain curve obtained
from this study is given in the Figure \ref{4.4.12}. It is seen in the figure that the elastic limit is at the strain value of 0.09. Beyond the elastic limit, the stress--strain curve departs from a straight line. Hence, its shape is permanently changed. The breaking point is observed at the strain rate of 0.23. The Young's modulus is determined as the slope of the stress-strain curve. Our calculated value of the Young's modulus of the $(10 \times 10)$ Carbon nanotube is 0.311 TPa.

Theoretical tensile strength is defined as the maximum stress which may be applied to the material without perturbing its stability. It can be given as
\begin{equation}
\sigma_{th}={1 \over s} \left( {\partial E_{tot} \over \partial \varepsilon } \right)_{\varepsilon=\varepsilon_i}
\end{equation}
where $\varepsilon$ is the stress, $\varepsilon_i$ is the maximum stress in the system, and $s$ is the surface area of cross--section. Our calculated value is 4.92 GPa, which is larger than that of Carbon fibers (2.6 GPa) \cite{endo}, but less than the in--plane tensile strength of graphite (20 GPa) \cite{pierso}.

Another mechanical property of interest is the Poisson ratio, defined by
\begin{equation}
\nu=- {1 \over \varepsilon} \left ( {R-R_{eq} \over R_{eq}} \right)
\end{equation}
where R is the radius of the tube at the strain $\varepsilon$, and $R_{eq}$ is the equilibrium (zero strain) tube radius. The Poisson ratio measures how much the tube contracts (expands) radially when subject to a positive (negative) axis strain $\varepsilon$. The corresponding value found in this study is 0.287.

\begin{figure}
\vskip 18 cm \includegraphics{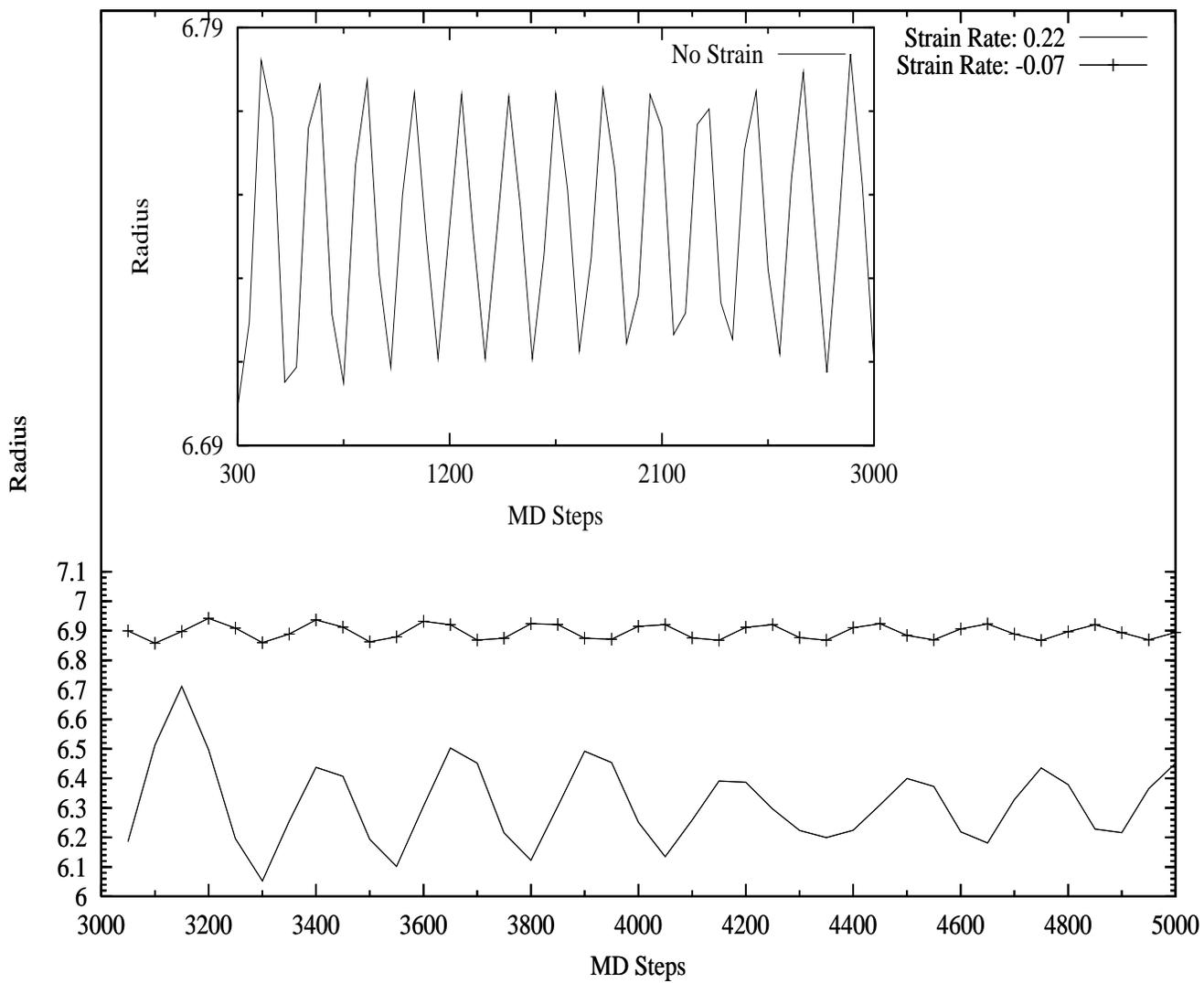} \vskip 1.5 cm
\caption{The variation of radius of (10x10) Carbon nanotube as a
function of MD steps with strain rates 22\% and -7\%;
respectively. It is also given for no strain case as inset. }
\label{4.4.13}
\end{figure}

Another interesting phenomenon we observed in simulation  is the
vibration of SWNT in radial direction. In Figure \ref{4.4.13}, the
average radius of the pristine $(10 \times 10)$ SWNT as a function
of MD steps is given. The frequency of vibration can be evaluated
from the figure and has the value of  $4.71\times 10^3$ GHz. The
variations of radius for the strains 0.22 and -0.07 are also given
in the Figure \ref{4.4.13}. It is found that the strain is
effective on the vibration frequency. Increasing strain on the
tube structure results in the decrease for the frequency of
vibration. It is ranged from $2.94\times10^3$ GHz to
$4.41\times10^3$ GHz with decrease by the increasing strain rate
and has a mean value of $3.70\times10^3$ GHz.

\medskip

We have thus determined  the elastic properties of a $(10 \times
10)$ carbon nanotube under tensile loading  and found the  Young's
modulus, tensile strength, Poisson ratio and frequency of
vibration to have the values 0.311 TPa, 4.92 GPa, 0.287 and $4.71
\times 10^3$ GHz, respectively. Several groups have reported a
wide range of values for the corresponding properties by using
various theoretical and experimental techniques.

The Young's modulus values given by different researchers range from 0.200 TPa to 5.5 TPa \cite{xizhma}. The following reasons may be given for this variety  of results:
\begin{itemize}
\item
The different values are used for the wall--thickness \cite{yabrbe}-\cite{lialdo}.
\item
Different procedures are applied to represent the strain
\cite{oziwmi},\cite{zhdugu},\cite{zhoshi},\cite{xizhma}.
\item
The curvature effect of nanotubes was neglected \cite{lu}, or not \cite{hegobe}. In Ref. \cite{lu}, it is concluded that the  elastic moduli of nanotubes, (SWNT and MWNT) were insensitive to geometrical structure while it is suggested in Ref. \cite{hegobe} that Young's modulus slightly depend on the tube diameter. On the other hand, the variation of Young's modulus as a function  of tube radius is also reported \cite{podoba,corwil}.
\item
Accuracy of methods: first--principle methods are more reliable
\cite{lialdo},\cite{oziwmi},\cite{zhdugu},\cite{zhoshi},\cite{xizhma}
 with comparison to the empirical potentials \cite{yazhch}- \cite{lu}.
\item
The strain rates that Young's modulus was calculated are either different \cite{xizhma} or not pointed out \cite{lu,oziwmi}.
\item
Difference in the tube lengths: although the periodic boundary condition is applied for the most cases, finite size effects might be still important.
\end{itemize}
Our result is in the range mentioned above. It emphasizes the high
Young's modulus and high strengths  of carbon nanotube. The strain
at tensile failure for SWNTs was predicted to be as high as 0.40
\cite{yabrbe}. This is a tensile strength of 400 GPa would be
expected for SWNTs if one used the in--plane Young's modulus of
graphite, $\sim$ 1 TPa \cite{ajaebb}. However, such a high tensile
strength has not been justified by experiments. In this study, it
is found that the elastic limit is at the strain rate 0.09 and
beyond this point tube becomes permanently changed. In Ref.
\cite{xizhma}, it is reported that for strain values greater than
0.10, the tube becomes softened. They also estimated the strain at
failure for the SWNT as 0.17 whereas it is found as 0.23 in this
study. The procedure for describing the strain in their work is to
apply hydrostatic stress to the tube wall. On the other hand,
Tight--Binding electronic calculations reported by Ozaki
{\itshape{et al.}\/} \cite{oziwmi} revealed a strain as high as
0.30.

The calculated and measured tensile strength also varies in value. In Ref. \cite{zhdugu}, it is reported as 6.249 GPa by a result of a first--principle study while it has the value of 62.9 GPa for the perfect 5x5 SWNT under hydrostatic pressure \cite{xizhma}. Another MD simulation by using the multi--body potential function of embedded atom method reports the tensile strength as 9.6 GPa \cite{yazhch}. On the other hand, it is reported as 3.6 GPa \cite{xilipa} and as ranged from 13 to 52 GPa \cite{yufiar} in the experimental studies. The value found in this study is 4.92 GPa and seems to comparable with the experimental and theoretical results.

The calculated Poisson ratio is 0.287 and in good agreement with the available reported values which are 0.278 \cite{lu}, and 0.32 \cite{zhdugu}. The evaluated frequency of vibration for the pristine $(10 \times 10)$ Carbon nanotube is $4.71\times10^3$ GHz which is very close to the value obtained from the experiment $4.94\times10^3$ GHz \cite{rariba} and almost same with the value reported in the MD simulation by using a bond--order potential \cite{zhoshi}. In Ref. \cite{zhoshi}, it is reported that the frequency of vibration is insensitive to the strain rate and the frequency of vibration is identified as self--vibration. We have found that it is not constant and increasing the strain rate decreases the vibration frequency.

\begin{acknowledgments}
We thank Dr. Tahir \c{C}a\u{g}{\i}n for
discussions and his help with $O(N)$ algorithms during  his TOKTEN/UNISTAR visit.
The research reported here is supported by T\"{U}B\.{I}TAK
(The Scientific and Technical Research Council of Turkey)
through the project TBAG-1877 and by the Middle East Technical University through the project AFP-2000-07-02-11.
\end{acknowledgments}

\end{document}